\documentclass[onecolumn,amsmath,amsmath,amssymb,11pt]{revtex4-1}

\usepackage{amsmath,amssymb,graphicx,bm,psfrag,bbold,float,color}

\usepackage[mathscr]{eucal}

\DeclareMathAlphabet{\mathpzc}{OT1}{pzc}{m}{it}

\begin{document}

\title{Quantum State Transmission in a Superconducting Charge Qubit-Atom Hybrid} 

\author{Deshui Yu$^{1}$, Mar\'ia Mart\'inez Valado$^{1}$, Christoph Hufnagel$^{1}$, Leong Chuan Kwek$^{1,2,3,4}$, Luigi Amico$^{1,5,6}$, \& Rainer Dumke$^{1,7}$}

\thanks{Correspondence and requests for material should be addressed to R.D. (email: rdumke@ntu.edu.sg).}

\affiliation{$^{1}$Centre for Quantum Technologies, National University of Singapore,
3 Science Drive 2, Singapore 117543, Singapore}

\affiliation{$^{2}$Institute of Advanced Studies, Nanyang Technological University,
60 Nanyang View, Singapore 639673, Singapore}

\affiliation{$^{3}$National Institute of Education, Nanyang Technological University,
1 Nanyang Walk, Singapore 637616, Singapore}

\affiliation{$^{4}$MajuLab, CNRS-UNS-NUS-NTU International Joint Research Unit, UMI 3654, Singapore}

\affiliation{\mbox{$^{5}$CNR-MATIS-IMM \& Dipartimento di Fisica e Astronomia, Universit\'a Catania, Via S. Soa 64, 95127 Catania, Italy}} 


\affiliation{$^{6}$INFN Laboratori Nazionali del Sud, Via Santa Sofia 62, I-95123 Catania, Italy}

\affiliation{$^{7}$Division of Physics and Applied Physics, Nanyang Technological University, 21 Nanyang Link, Singapore 637371, Singapore}

\maketitle

{\bf Hybrids consisting of macroscopic superconducting circuits and microscopic components, such as atoms and spins, have the potential of transmitting an arbitrary state between different quantum species, leading to the prospective of high-speed operation and long-time storage of quantum information. Here we propose a novel hybrid structure, where a neutral-atom qubit directly interfaces with a superconducting charge qubit, to implement the qubit-state transmission. The highly-excited Rydberg atom located inside the gate capacitor strongly affects the behavior of Cooper pairs in the box while the atom in the ground state hardly interferes with the superconducting device. In addition, the DC Stark shift of the atomic states significantly depends on the charge-qubit states. By means of the standard spectroscopic techniques and sweeping the gate voltage bias, we show how to transfer an arbitrary quantum state from the superconducting device to the atom and vice versa.}\\

A quantum computer makes direct use of qubits to encode information and perform operations on data according to the laws of quantum mechanics~\cite{Book:Nielsen2000}. Due to the properties of superposition and entanglement of quantum states, such a computing device is expected to operate exponentially faster than a classical computer for certain problems. Recently, some basic quantum logic gates have been executed on various quantum systems composed of a small number of qubits, for instances, trapped ions~\cite{Nature:Kielpinski2002}, neutral atoms~\cite{PRL:Brennen1999}, photons~\cite{RMP:Kok2007}, NMR~\cite{Nature:Vandersypen2011}, and superconducting (SC) circuits~\cite{RMP:Makhlin2001}. However, the development of an actual quantum computer is still in its infancy since no quantum system practically fulfills all DiVincenzo criteria~\cite{PRA:Loss1998} for the physical implementation of quantum computation.

Hybridizing different quantum systems could inherit the advantages of each component and compensate the weaknesses with each other~\cite{PRL:Tian2004,PRL:Rabl2006,RMP:Xiang2013}. A promising structure is to combine the SC circuits with neutral atoms. Macroscopic solid-state devices including submicrometer-sized Josephson junctions (JJ) possess the advantages of rapid information processing ($\sim1$ ns), flexibility, and scalability. However, due to the strong coupling to the local electromagnetic environment, the relaxation and dephasing times of the SC circuits, which are of the order of $10~\textrm{ns}\sim100~\mu$s~\cite{Science:Mooij2003,PRL:Nakamura2002,PRB:Duty2004,PRA:Koch2007,PRB:Rigetti2012}, are significantly limited by the $1/f$ fluctuations in background charge, flux, and critical current~\cite{Science:Vion2002} and even the readout back-action~\cite{PRB:Schreier2008}. In contrast, the microscopic atomic systems are characterized by precise quantum-state control and long coherence time ($1~\textrm{ms}\sim1~\textrm{s}$), though they own a relatively long gate operation time because of the weak coupling to external fields~\cite{PRL:Monroe1995,Nat:Monroe2002,PRL:Isenhower2010,RMP:Saffman2010} and have limited scalability. Transmitting information between these two distinct quantum realizations could lead to the rapid processing and long-term storage of quantum states, where the SC circuits serve as the fast processor while the atoms play the role of memory~\cite{AnnuRevCondensMatterPhys:Daniilidis2013,PNAS:Kurizkia2015,PRA:Petrosyan2009}.

The SC circuits and atoms can be indirectly coupled by integrating both of them on a microwave SC cavity, such as a LC resonator or a coplanar waveguide (CPW) resonator, which acts as a data bus to transfer the quantum information between the atomic memory and the SC processor~\cite{PRL:Maitre1997,RMP:Raimond2001,Nature:Wallraff2004,Nature:Majer2007}. However, the large detunings of the off-resonance SC qubit-resonator and atom-resonator interactions significantly weaken the virtual-photon-mediated SC qubit-atom coupling. Moreover, the fluctuation of intraresonator photon number increases the dephasing rate of qubits~\cite{PRB:Rigetti2012}.

The atoms can also directly talk to the SC devices via interacting with the local electromagnetic field. The current relevant research mainly focuses on the information transmission between neutral atoms and flux qubits, where the low-lying atomic states couple to the microwave-frequency alternating magnetic field from SC loops~\cite{RevMexFis:Hoffman2011,PRA:Patton2013}. Although replacing a single atom by an ensemble of $N$ atoms can enhance the magnetic intersubsystem coupling by a factor of $\sqrt{N}$, the atomic number fluctuation and the interparticle interaction challenge the experimental implementation. These issues may be solved by employing the electric dipole interface between the highly-excited Rydberg atomic states and local electric field from SC devices~\cite{PRA:Yu2016}.

Here, we propose a hybrid scheme, where a charge qubit is electrically coupled to an atomic qubit comprised of the ground and Rydberg states. The neutral atom placed inside the gate capacitor acts as the dielectric medium and affects the gate capacitance, resulting in the modulated charge-qubit energy bands. In addition, the local quasi-static electric field strongly depends on the charge-qubit state, leading to different DC Stark shifts of atomic-qubit states. We show that an arbitrary quantum state can be transmitted between these two distinct qubits. The two-qubit controlled-NOT (CNOT) logic gate and single-qubit Hadamard transform, which are necessary to entangle two qubits with different species and induce a $\pi$-rotation of the control qubit, respectively, in the state-transmission protocol, can be implemented by means of standard spectroscopic techniques and sweeping the gate charge bias. Our state-transmission protocol also provides a potential for transferring the quantum state between or remotely entangling two distant noninteracting SC qubits via the flying-qubit-linked atoms.\\

{\bf RESULTS}\\

\begin{figure*}
\includegraphics[width=8cm]{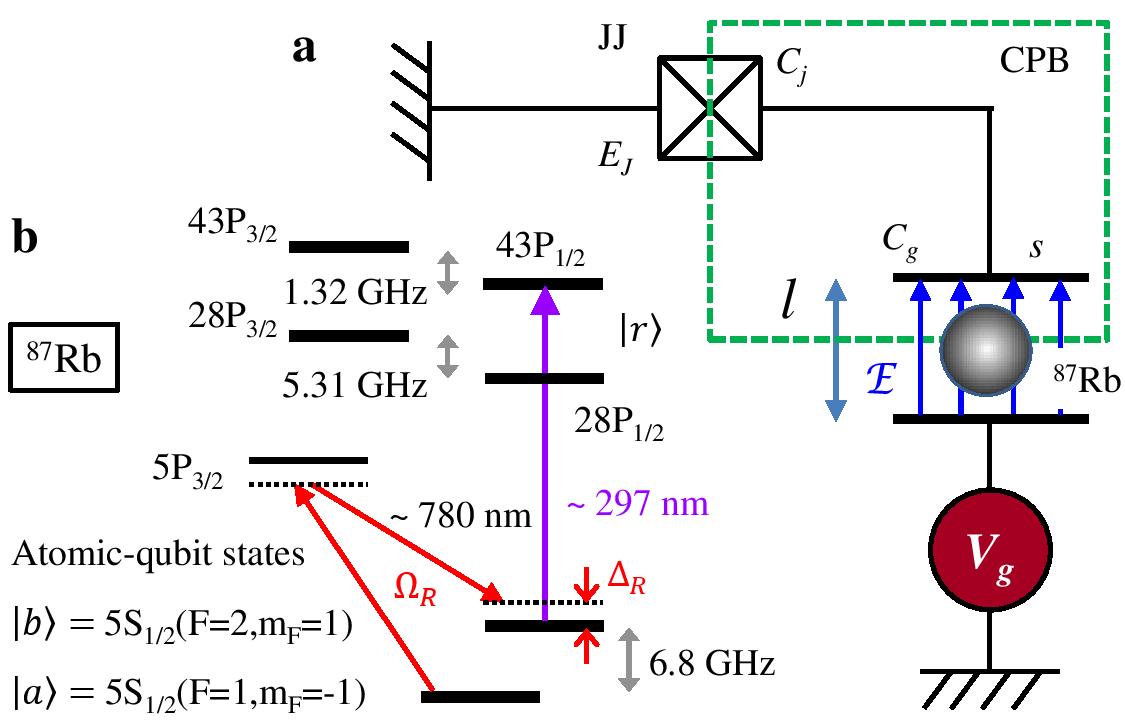}\\
\caption{{\bf Hybrid quantum circuit.} (a) A voltage source $V_{g}$ drives a single CPB via a parallel-plate capacitor $C_{g}$ with the plate area $s$ and the interplate distance $l$. Cooper pairs tunnel back and forth between CPB and the SC reservoir through a JJ with a Josephson coupling energy $E_{J}$ and a self-capacitance $C_{j}$. A $^{87}$Rb atom located inside $C_{g}$ plays the role of dielectric medium and interfaces with the local electric field ${\mathpzc{E}}$. (b) Schematic atomic level structure. Two hyperfine ground $|a\rangle=5S_{1/2}(F=1,m_{F}=-1)$ and $|b\rangle=5S_{1/2}(F=2,m_{F}=1)$ states comprise the atomic qubit. The qubit transition is implemented via the Raman transition with the intermediate $5P_{3/2}$ state. A highly-excited Rydberg $|r\rangle=nP_{1/2}$ state acting an auxiliary role is employed to enhance the SC circuit-atom interaction. The relatively small hyperfine splitting of $|r\rangle$ is neglected. The atom in $|u=a,b\rangle$ can be directly excited to $|r\rangle$ via a resonant light pulse at 297 nm. The microwave-frequency alternating electric field ${\mathpzc{E}}$ does not induce any optical atomic transitions associated with $|u=a,b,r\rangle$.}\label{Figure-1}
\end{figure*}

{\bf Charge qubit-atom Hybrid.} We consider a simple SC charge qubit~\cite{PRL:Nakamura1997,PhysScripta:Bouchiat1998,Nature:Nakamura1999}, where a single Cooper pair box (CPB) is connected to a SC reservoir via a JJ with a low self-capacitance $C_{j}$ (see Fig.~\ref{Figure-1}a). The Cooper pairs can tunnel into or out of the box at a rate of $E_{J}=\frac{\Phi_{0}I_{c}}{2\pi}$ (the magnetic flux quantum $\Phi_{0}$ and the critical current $I_{c}$ of JJ). The CPB is biased by a voltage source $V_{g}$ via a parallel-plate capacitor $C_{g}$ with the plate area $s$ and the interplate separation $l$. The Hamiltonian describing the dynamics of excess Cooper pairs in the box is written as
\begin{equation}
H_{c}=E_{C}\left(N-N_{g}\right)^{2}-\frac{E_{J}}{2}\left(e^{i\delta}+e^{-i\delta}\right),
\end{equation}
where $E_{C}=\frac{(2e)^{2}}{2(C_{g}+C_{j})}\gg E_{J}$ gives the Coulomb charging energy, $N_{g}=\frac{C_{g}V_{g}}{2e}$ is the offset charge, and $\delta$ is the phase drop across the JJ. The operator $N$ counts the number of excess Cooper pairs in the box, $N|\tilde{n}\rangle=\tilde{n}|\tilde{n}\rangle$. Around the charge-degenerate spot ($N_{g}= 0.5$), two lowest charge states $|\tilde{n}\rangle = 0$ and $|\tilde{n}\rangle=1$ are well separated from others and implement the charge qubit. We have omitted the work done by the gate voltage, whose effect on the system can be neglected.

A $^{87}$Rb atom placed inside the gate capacitor $C_{g}$ interacts with the internal electric field ${\mathpzc{E}}$ and plays a role of dielectric medium. The direction of ${\mathpzc{E}}$ is chosen as the the quantization axis. Two hyperfine ground states $|a\rangle=5S_{1/2}(F=1,m_{F}=-1)$ and $|b\rangle=5S_{1/2}(F=2,m_{F}=1)$ with an energy spacing of 6.8 GHz are applied to form an atomic qubit (see Fig.~\ref{Figure-1}b), where $F$ is the total angular momentum quantum number and $m_{F}$ gives the corresponding projection along the $z$-axis. The qubit-state flipping of the atom is achieved by the resonant Raman transition via the intermediate $5P_{3/2}$ state. A highly-excited Rydberg state $|r\rangle=nP_{1/2}$ is employed as an auxiliary state to enhance the charge qubit-atom interaction. In comparison with $E_{J}$ and $E_{C}$, the hyperfine splitting of $|r\rangle$, which is of the order of several MHz~\cite{PRA:Tauschinsky2013}, can be neglected. Here $n$ denotes the principle quantum number of Rydberg atom. A resonant $\pi$-laser pulse at 297 nm transfers the atomic component in $|a\rangle$ or $|b\rangle$ completely to $|r\rangle$.

The Cooper-pair tunneling through the JJ varies the internal electric field ${\mathpzc{E}}$ with a frequency typically of the order of $E_{J}/\hbar$ (the reduced Planck's constant $\hbar$). The energy spacings of any electric-dipole transitions associated with $|r\rangle$ are much larger than $E_{J}$. Thus, ${\mathpzc{E}}$ can be treated as quasi-static. In the weak-field limit, the capacitance $C_{g}$ with the atom in $|u=a,b,r\rangle$ is expressed as $C^{(u)}_{g}=C_{g0}+\frac{s}{l}\frac{\alpha_{u}}{V}$, where $C_{g0}=\frac{\epsilon_{0}s}{l}$ gives the empty gate capacitance (without the atom), $\epsilon_{0}$ is the vacuum permittivity, $\alpha_{u}$ denotes the static polarizability of the atom in $|u\rangle$, and $V=sl$ is the volume of homogeneous atomic distribution over the gate capacitor. 

According to $C^{(u)}_{g}$, the Coulomb energy $E_{C}$, the offset charge $N_{g}$, and the internal electric field ${\mathpzc{E}}$ are rewritten as $E^{(u)}_{C}=\frac{E_{C0}}{1+\eta_{u}}$, $N^{(u)}_{g}=N_{g0}+\eta_{u}(N_{g0}+N_{j})$, and ${\mathpzc{E}}_{u}={\mathpzc{E}}_{0}\frac{N+N_{j}}{1+\eta_{u}}$~\cite{JLowTempPhys:Pekola2012}, respectively, where the empty charging energy $E_{C0}=\frac{(2e)^{2}}{2(C_{g0}+C_{j})}$, the empty gate charge bias $N_{g0}=\frac{C_{g0}V_{g}}{2e}$, $N_{j}=\frac{C_{j}V_{g}}{2e}$, and the field amplitude ${\mathpzc{E}}_{0}=\frac{1}{l}\frac{2e}{C_{g0}+C_{j}}$. The ratio $\eta_{u}=\frac{1}{C_{g0}+C_{j}}\frac{s}{l}\frac{\alpha_{u}}{V}$ measures the relative variation of the total box capacitance caused by the single atom. A large $\eta_{u}$ reduces $E^{(u)}_{C}$ but enhances $N^{(u)}_{g}$. Combining the energy associated with the atom, the system Hamiltonian is given by
\begin{equation}
H=\sum_{u=a,b,r}(H^{(u)}_{c}+\hbar\omega_{u}+\Delta E_{u})P_{u},
\end{equation}
where $H^{(u)}_{c}$ denotes the SC-circuit Hamiltonian $H_{c}$ with the atom in $|u\rangle$, $\hbar\omega_{u}$ is the intrinsic atomic energy of $|u\rangle$, $\Delta E_{u}=-\frac{1}{2}\alpha_{u}{\mathpzc{E}}^{2}_{u}$ indicates the DC Stark shift of $|u\rangle$ induced by the electric field ${\mathpzc{E}}_{u}$, and $P_{u}=|u\rangle\langle u|$ is the atomic projection operator. We restrict ourselves within the Hilbert space spanned by $\{|u,\tilde{n}\rangle=|u\rangle\otimes|\tilde{n}\rangle,u=a,b,r;\tilde{n}=0,1\}$ and obtain a hybrid system consisting of a charge qubit ($|0\rangle$ and $|1\rangle$) and an atomic qubit ($|a\rangle$ and $|b\rangle$). The auxiliary Rydberg $|r\rangle$ state enables the strong interface between SC circuit and atom. Diagonalizing $H$ gives us the eigenvalues ${\mathcal{E}}^{(u)}_{k}$ and eigenstates $\psi^{(u)}_{k}$ of the hybrid system,
\begin{equation}
H\psi^{(u)}_{k}={\mathcal{E}}^{(u)}_{k}\psi^{(u)}_{k},
\end{equation}
with $k=0,1$ denoting the different energy bands for a given $|u\rangle$.

For the zero gate voltage $V_{g}=0$, we have $N_{g0}=N_{j}=0$ and the hybrid-system eigenenergies are analytically derived as
\begin{eqnarray}\label{Eigenenergies}
{\mathcal{E}}^{(u=a,b,r)}_{k=0,1}&=&\hbar\omega_{u}+\frac{1}{2}(E^{(u)}_{C}+\Delta E_{u})-(-1)^{k}\frac{1}{2}\sqrt{(E^{(u)}_{C}+\Delta E_{u})^{2}+E^{2}_{J}},
\end{eqnarray}
and the corresponding eigenstates are given by
\begin{equation}\label{Eigenstates}
\psi^{(u=a,b,r)}_{k=0,1}=\frac{E_{J}|u,0\rangle+({\mathcal{E}}^{(u)}_{k}-\hbar\omega_{u})|u,1\rangle}{\sqrt{E^{2}_{J}+({\mathcal{E}}^{(u)}_{k}-\hbar\omega_{u})^{2}}}.
\end{equation}
In the limit of $E^{(u)}_{C}\gg E_{J}$, we are left with ${\mathcal{E}}^{(u)}_{0}\simeq\hbar\omega_{u}$, ${\mathcal{E}}^{(u)}_{1}\simeq\hbar\omega_{u}+E^{(u)}_{C}$, $\psi^{(u)}_{0}\simeq|u,0\rangle$, and $\psi^{(u)}_{1}\simeq|u,1\rangle$. For the atom in the hyperfine ground states $|u=a,b\rangle$, whose static polarizabilities $\alpha_{u}=0.079$ Hz/(V/cm)$^{2}$~\cite{Book:Miller2000} are extremely small, we obtain $\eta_{u}\simeq0$, $E^{(u)}_{C}\simeq E_{C0}$, $N^{(u)}_{g}\simeq N_{g0}$, and $\Delta E_{u}\simeq0$, meaning the atom hardly affects the SC circuit. Thus, the energy difference between $|r,0\rangle$ and $|a,0\rangle$ approximates to the intrinsic energy spacing $\hbar\omega_{ra}=\hbar\omega_{r}-\hbar\omega_{a}$, i.e., ${\mathcal{E}}^{(r)}_{0}-{\mathcal{E}}^{(a)}_{0}\simeq\hbar\omega_{ra}$, while the energy separation between $|r,1\rangle$ and $|a,1\rangle$ is shifted away from $\hbar\omega_{ra}$, i.e., ${\mathcal{E}}^{(r)}_{1}-{\mathcal{E}}^{(a)}_{1}\simeq\hbar\omega_{ra}+(E^{(r)}_{C}+\Delta E_{r}-E_{C0})$. Similarly, we have ${\mathcal{E}}^{(r)}_{0}-{\mathcal{E}}^{(b)}_{0}\simeq\hbar\omega_{rb}$ and ${\mathcal{E}}^{(r)}_{1}-{\mathcal{E}}^{(b)}_{1}\simeq\hbar\omega_{rb}+(E^{(r)}_{C}+\Delta E_{r}-E_{C0})$.

Distinguishing the system energy spectrum with the atom in $|r\rangle$ from that associated with $|u=a,b\rangle$, a large $\eta_{r}$ is necessary to induce the apparent variations of $E^{(r)}_{C}$ and $N^{(r)}_{g}$ compared with the small Josephson energy $E_{J}$ and the ratio $E_{J}/E_{C0}$, respectively, as well as a strong DC Stark shift $\Delta E_{r}$. Thus, the SC circuit should be carefully designed and the Rydberg $|r\rangle$ state needs to be chosen accordingly. As a specification, we list the structure of CPB in Table~\ref{Table}.

When a Rydberg atom is brought into the vicinity of SC circuit, the inhomogeneous stray electric fields originating from the contaminations on the cryogenic surface are particular detrimental to the quantum hybrid system since they cause the unwanted energy-level shifts and destroy the atomic coherence. However, there might be ways to mitigate or circumvent the effects of stray electric fields. It has been shown that the direction of electric field produced by the adsorbates due to the chemisorption or physisorption depends on the material properties~\cite{PRL:Chan2014}. In principle, one can envision to pattern the surfaces with two materials which give rise to opposing dipole moments of adsorbates. Furthermore, as experimentally demonstrated in~\cite{RPA:HermannAvigliano2014}, the stray fields can be minimized by saturating the adsorbates film. The remaining uniform electric fields could be canceled by applying offset fields.

The extra measures of reducing the effects of stray fields possibly affect the performance of hybrid system in a different manner. One way to estimate the dependence of the coherence time of a qubit on the surface properties is to investigate the surface-dependent change of the $Q$-factor of a cavity. Only a few studies have been done so far investigating the superconducting cavity for various materials absorbed to the surface~\cite{Knobloch:1998zt}. With the knowledge at hand, it is hard to estimate the effect of a physisorbed layer of rubidium or specific protective coatings on the superconducting system. Here we assume the resulting decoherence time of the atomic qubit close to the surface is much longer than that of the charge qubit. In the following, we discuss the quantum-state transfer between two different qubits.

{\bf State transmission from atom to SC circuit.} Transferring an arbitrary qubit state from the atom to the charge qubit primarily relies on a two-qubit CNOT logic gate, where the state flipping of the charge qubit is conditioned on the atomic-qubit state, and a one-qubit Hadamard gate acting on the atom~\cite{Book:Nielsen2000}. For performing the CNOT operation, the polarizability $\alpha_{r}$ of Rydberg $|r\rangle$ state should be large enough that the atom is strongly coupled with the SC circuit. In addition, the corresponding internal electric field ${\mathpzc{E}}_{r}$ needs to be smaller than the first avoided crossing field of $|r\rangle$~\cite{PRA:Sullivan1985}. Based on the specification of CPB structure listed in Table~\ref{Table}, we choose $|r\rangle=43P_{1/2}$, whose relevant physical parameters are derived from~\cite{Thesis:Pritchard,JPB:Low2012,JPB:Branden2010} and also shown in Table~\ref{Table}.

\begin{figure*}
\includegraphics[width=16cm]{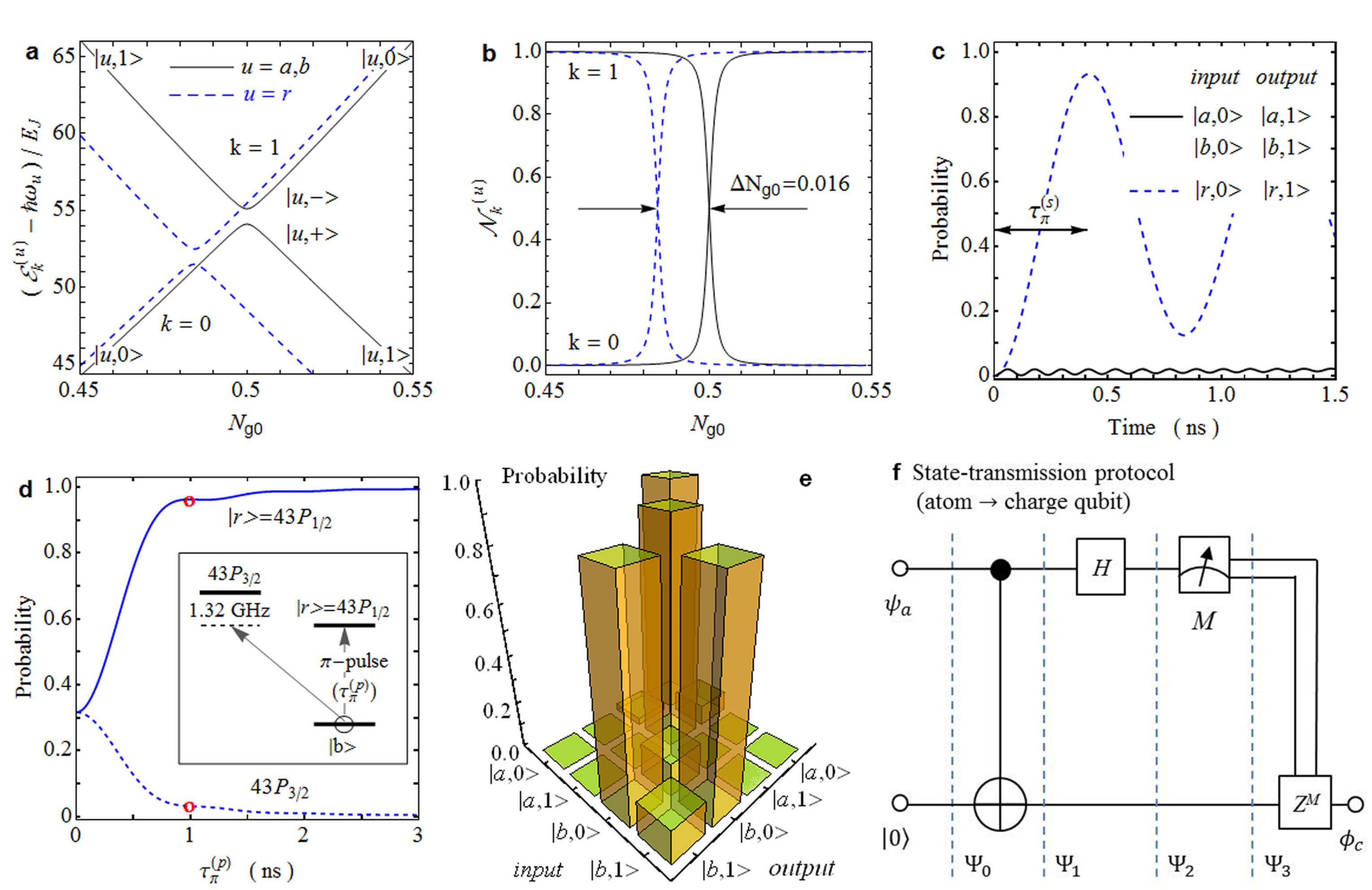}\\
\caption{{\bf Quantum-state transmission from atom to charge qubit.} (a) Eigenenergy bands ${\mathcal{E}}^{(u=a,b,r)}_{k=0,1}$, which are shifted by the atomic-state energies $\hbar\omega_{u}$, as a function of the empty charge bias $N_{g0}$ with $|r\rangle=43P_{1/2}$. The atom playing a dielectric role changes the gate capacitance $C_{g}$, resulting in the atomic-state-dependence energies of CPB. The corresponding expectation values ${\mathcal{N}}^{(u)}_{k=0,1}$ of numbers of excess Cooper pairs inside the box are shown in (b). Two charge-degeneracy points with the atom respectively in $|u=a,b\rangle$ and $|r\rangle$ stand $\Delta N_{g0}=0.016$ apart. (c) Time-dependent probability for the relaxing hybrid system in an $output$ state, given the system initially prepared in an $input$ state. The empty gate charge bias is set at $N_{g0}=0.5-\Delta N_{g0}$. At the $\pi$-pulse duration $\tau^{(s)}_{\pi}=\pi\hbar/E_{J}$, the probability for the system in $output=|r,1\rangle$ is 0.93 with $input=|r,0\rangle$. (d) Atomic populations in $|r\rangle$ and $43P_{3/2}$ vs. the $\pi$-light pulse length $\tau^{(p)}_{\pi}$ with the atom initially prepared in $|b\rangle$. For the three-level system composed of $|b\rangle$, $|r\rangle$, and $43P_{3/2}$, the $\pi$-light pulse is resonantly coupled to the $|b\rangle-|r\rangle$ transition. The atom population in $|r\rangle$ reaches 0.96 at $\tau^{(p)}_{\pi}=1.0$ ns. (e) Truth table amplitudes of the CNOT gate with taking into account the relaxation and dephasing of the charge qubit. The standard process fidelity is ${\cal{F}}=0.83$. (f) Scheme of transmitting an arbitrary atomic-qubit state $\psi_{a}=\mu|a\rangle+\nu|b\rangle$ to the charge qubit, which is initially prepared in $|0\rangle$, via the Hadamard gate $H$, phase-flip gate $Z$, and measurement $M$. $\Psi_{0-3}$ are the intermediate two-qubit states. The final charge qubit is $\phi_{c}=\mu|0\rangle+\nu|1\rangle$.}\label{Figure-2}
\end{figure*}

We first consider the system energy spectrum. Figure~\ref{Figure-2}a illustrates the shifted eigenenergies $({\mathcal{E}}^{(u)}_{k=0,1}-\hbar\omega_{u})$ with $u=a,b,r$ versus the empty charge bias $N_{g0}$ around $N_{g0}=0.5$. For $u=a$ and $b$, $({\mathcal{E}}^{(u)}_{k=0,1}-\hbar\omega_{u})$ are nearly same to that of a common charge qubit due to $\eta_{u}\simeq0$. An avoided energy-level crossing occurs at $N_{g0}=0.5$, where ${\mathcal{E}}^{(u)}_{0}$ and ${\mathcal{E}}^{(u)}_{1}$ approach each other with a minimal energy spacing of $E_{J}$. The Cooper-pair tunneling takes effect only around the charge-degenerate point $N_{g0}=0.5$ within a narrow region $|N_{g0}-0.5|<E_{J}/E_{C0}$~\cite{Vion2004}. In contrast, the energy bands $({\mathcal{E}}^{(r)}_{k=0,1}-\hbar\omega_{r})$ move down relative to $({\mathcal{E}}^{(u)}_{k=0,1}-\hbar\omega_{u})$ with $u=a,b$ due to the enhanced $E^{(r)}_{C}$ and large DC Stark shift $\Delta E_{r}$. The minimal separation between ${\mathcal{E}}^{(r)}_{0}$ and ${\mathcal{E}}^{(r)}_{1}$, however, is still determined by the Josephson energy $E_{J}$. The position of the corresponding energy-level anticrossing shifts to the left side of $N_{g0}=0.5$ because of the enlarged offset charge $N^{(r)}_{g}$. At either avoided crossing, the hybrid system stays in the superposition states $\psi^{(u)}_{k=0}=|u,+\rangle$ and $\psi^{(u)}_{k=1}=|u,-\rangle$ with $|u,\pm\rangle=(|u,1\rangle\pm|u,0\rangle)/\sqrt{2}$. We also show the expectation values of excess Cooper-pair numbers, ${\mathcal{N}}^{(u=a,b,r)}_{k=0,1}=\langle\psi^{(u)}_{k}|N|\psi^{(u)}_{k}\rangle$, in Fig.~\ref{Figure-2}b and find that two charge-degenerate spots are separated by $\Delta N_{g0}=0.016$ larger than the ratio $E_{J}/E_{C0}=0.004$, indicating the shifted energy spectra with the atom in $|u=a,b\rangle$ and $|r\rangle$ can be well distinguished.

The dependence of the avoided-level crossing on the atomic state allows us to control the charge-qubit transition via preparing the atom in different states. Setting the empty charge bias at $N_{g0}=0.5-\Delta N_{g0}$, the hybrid system resonantly oscillates between $|r,0\rangle$ and $|r,1\rangle$ with a half period ($\pi$-pulse duration) of $\tau^{(s)}_{\pi}=\pi\hbar/E_{J}=0.4$ ns while the $|u,0\rangle-|u,1\rangle$ transitions with $u=a,b$ are strongly suppressed due to the large detuning as shown in Fig.~\ref{Figure-2}c, where the master equation involving the relaxation and dephasing of charge qubit~\cite{PRA:Boissonneault2009} is employed. It is seen that the probability of the system switching between $|r,0\rangle$ and $|r,1\rangle$ reaches 0.93 at $\tau^{(s)}_{\pi}$.

The two-qubit CNOT gate, where the atom acts as the control qubit while the charge qubit plays the target role, can be implemented via three steps: (1) Initially, the gate voltage stays at zero, $V_{g}=N_{g0}=0$. Two $\pi$-light pulses (the time duration $\tau^{(p)}_{\pi}$) at 297 nm are applied resonantly on the $|b,0\rangle-|r,0\rangle$ and $|b,1\rangle-|r,1\rangle$ transitions to transfer the populations in $|b,0\rangle$ and $|b,1\rangle$ completely to $|r,0\rangle$ and $|r,1\rangle$, respectively. Thus, different components $|a,\tilde{n}=0,1\rangle$ and $|r,\tilde{n}=0,1\rangle$ are spectroscopically discriminated (Fig.~\ref{Figure-2}a). (2) The empty charge bias $N_{g0}$ nonadiabatically raises to the charge-degenerate point for two adiabatic energy curves associated with $|r,0\rangle$ and $|r,1\rangle$, i.e., $N_{g0}=(0.5-\Delta N_{g0})$. After staying at this sweet spot for the $\pi$-pulse duration of $\tau^{(s)}_{\pi}$, $N_{g0}$ decreases back to zero nonadiabatically. As a result, the populations in $|r,0\rangle$ and $|r,1\rangle$ switches with each other while that in $|a,0\rangle$ and $|a,1\rangle$ do not change. (3) The $\pi$-light pulses are used again to bring the populations in $|r,0\rangle$ and $|r,1\rangle$ back to $|b,0\rangle$ and $|b,1\rangle$, respectively, without affecting the components of $|a,0\rangle$ and $|a,1\rangle$.

In steps (1) and (3), the intensities of the light pulses need to be strong enough to reduce the pulse duration $\tau^{(p)}_{\pi}$ shorter than the decoherence time of the charge qubit. The experimentally feasible light-pulse length can be as small as 1 ns with a corresponding effective Rabi frequency of the order of 1 GHz~\cite{PRL:Huber2011}. However, for such a strong atom-light interaction, the small fine-structure splitting between $|r\rangle=43P_{1/2}$ and $43P_{3/2}$, i.e., $2\pi\times1.32$ GHz~\cite{PRA:Li2003}, affects the atom transfer between $|b\rangle$ and $|r\rangle$. To suppress the unwanted population in $43P_{3/2}$, the $\pi$-pulse length $\tau^{(p)}_{\pi}$ should be chosen to fit the experimental conditions. According to Fig.~\ref{Figure-2}d, $\tau^{(p)}_{\pi}$ can be set at 1 ns, much shorter than the decoherence time of the charge qubit, with the corresponding Rabi frequency of $2\pi\times0.5$ GHz. The resulting atom-transfer efficiency is over 0.96. Due to the large ground-state hyperfine splitting of $2\pi\times6.8$ GHz, the light pulses hardly affect the components of $|a,0\rangle$ and $|a,1\rangle$.

We numerically simulate the CNOT operation with different $input$ and $output$ states via applying the master equation~\cite{PRA:Boissonneault2009}. The whole gate operation duration is less that 2.5 ns. The resulting register populations are depicted in Fig.~\ref{Figure-2}e based on the specification. It is seen that the quantum logic gate preserves the charge-qubit states when the atom is prepared in $|a\rangle$, whereas for the atom in $|b\rangle$ the charge-qubit state switches between $|0\rangle$ and $|1\rangle$ with high probabilities. The standard process fidelity~\cite{PRA:Gilchrist2005,PRA:Bongioanni2010} reaches ${\cal{F}}=0.83$.

After performing the CNOT gate, the transmission of quantum state from the atom to SC circuit is straightforward. We assume that the atomic qubit is initially in an arbitrary state $\psi_{a}=\mu|a\rangle+\nu|b\rangle$ while the charge qubit is prepared in $|0\rangle$, leading to the system state $\Psi_{0}=\psi_{a}\bigotimes|0\rangle=\mu|a,0\rangle+\nu|b,0\rangle$ (see Fig.~\ref{Figure-2}e). Passing through the CNOT gate, the system state becomes $\Psi_{1}=\mu|a,0\rangle+\nu|b,1\rangle$. Then, a single-qubit Hadamard gate acts on the atom and the hybrid system arrives at $\Psi_{2}=\frac{1}{\sqrt{2}}|a\rangle\bigotimes(\mu|0\rangle+\nu|1\rangle)+\frac{1}{\sqrt{2}}|b\rangle\bigotimes(\mu|0\rangle-\nu|1\rangle)$. Afterwards, we measure the atomic-qubit state and obtain $\Psi_{3}=|a\rangle\bigotimes(\mu|0\rangle+\nu|1\rangle)$ or $\Psi_{3}=|b\rangle\bigotimes(\mu|0\rangle-\nu|1\rangle)$ depending on the readout which triggers an extra Pauli-$Z$ (phase-flip) gate~\cite{JPA:Obada2012} acting on the SC device. As a result, the final charge qubit is in $\phi_{c}=\mu|0\rangle+\nu|1\rangle$ and the quantum-state transmission is accomplished.

The Hadamard gate for atomic qubit can be performed via the pulsed two-photon $|a\rangle-5P_{3/2}-|b\rangle$ Raman process with a Raman detuning $\Delta_{R}$ and a Raman coupling strength $\Omega_{R}$ (see Fig.~\ref{Figure-1}b). After adiabatically eliminating the $5P_{3/2}$ state, one obtains an effective light-driven two-state ($|a\rangle$ and $|b\rangle$) system with the detuning $\Delta_{R}$ and the Rabi frequency $\Omega_{R}$. Choosing $\Delta_{R}=\Omega_{R}>0$ and the light-pulse length $\tau_{R}=\pi/(\sqrt{2}\Omega_{R})$ leads to the time evolution operator of the atom $U(\tau_{R})=\frac{1}{\sqrt{2}}(|a\rangle\langle a|+|b\rangle\langle b|-|a\rangle\langle b|+|b\rangle\langle a|)$. The operator $U(\tau_{R})$ maps the atomic states $|a\rangle$ and $|b\rangle$ onto $\frac{1}{\sqrt{2}}(|a\rangle+|b\rangle)$ and $\frac{1}{\sqrt{2}}(|a\rangle-|b\rangle)$, respectively, achieving the Hadamard transformation.

{\bf State transmission from SC circuit to atom.} Similarly, the protocol for transferring an arbitrary charge-qubit state to the atom relies on a two-qubit CNOT gate, where the atom flips its state conditioned on the charge-qubit state, and a one-qubit Hadamard gate acting on the SC circuit. We first consider the CNOT operation. The gate voltage $V_{g}$ is set at zero, resulting in $N_{g0}=N_{j}=0$. A large atomic polarizability $\alpha_{r}$ for $|r\rangle$, which leads to a strong DC Stark shift $\Delta E_{r}$, is necessary for spectroscopic distinguishing four $|a,0\rangle-|r,0\rangle$, $|a,1\rangle-|r,1\rangle$, $|b,0\rangle-|r,0\rangle$, and $|b,1\rangle-|r,1\rangle$ transitions. However, the single-qubit Hadamard operation on the charge qubit, which is performed via adiabatically sweeping $N_{g0}$ from 0 to 0.5, requires that the charge-degenerate spots with the atom in different states approximately locate at $N_{g0}=0.5$, i.e., $\Delta N_{g0}\ll E_{J}/E_{C0}$. Hence, $\alpha_{r}$ should not be very large. As an example, we employ $|r\rangle=28P_{1/2}$, whose relevant physical parameters are listed in Table~\ref{Table}. The corresponding $\eta_{r}$ approximates zero, indicating the very weak effect of the atom in $|r\rangle$ on the SC circuit, and the avoided crossing between $|r,0\rangle$ and $|r,1\rangle$ occurs at $N_{g0}=0.5$. Moreover, at $N_{g0}=0$ the energy spacings of different transitions are ${\mathcal{E}}^{(r)}_{0}-{\mathcal{E}}^{(a)}_{0}\simeq\hbar\omega_{ra}$, ${\mathcal{E}}^{(r)}_{1}-{\mathcal{E}}^{(a)}_{1}\simeq\hbar\omega_{ra}+\Delta E_{r}$, ${\mathcal{E}}^{(r)}_{0}-{\mathcal{E}}^{(b)}_{0}\simeq\hbar\omega_{rb}$, and ${\mathcal{E}}^{(r)}_{1}-{\mathcal{E}}^{(b)}_{1}\simeq\hbar\omega_{rb}+\Delta E_{r}$. For our physical specification, the DC Stark shift $\Delta E_{r}$ of $|r\rangle$ reaches $2\pi\times0.2$ GHz. A $\pi$-laser pulse resonant to the $|a,1\rangle-|r,1\rangle$ transition switches the atomic state between $|a\rangle$ and $|r\rangle$ and keeps the charge qubit in $|1\rangle$, but this pulse weakly interacts with the $|a,0\rangle-|r,0\rangle$ transition due to the detuning $\Delta E_{r}$, which is applicable to the CNOT operation.

\begin{figure*}
\includegraphics[width=16cm]{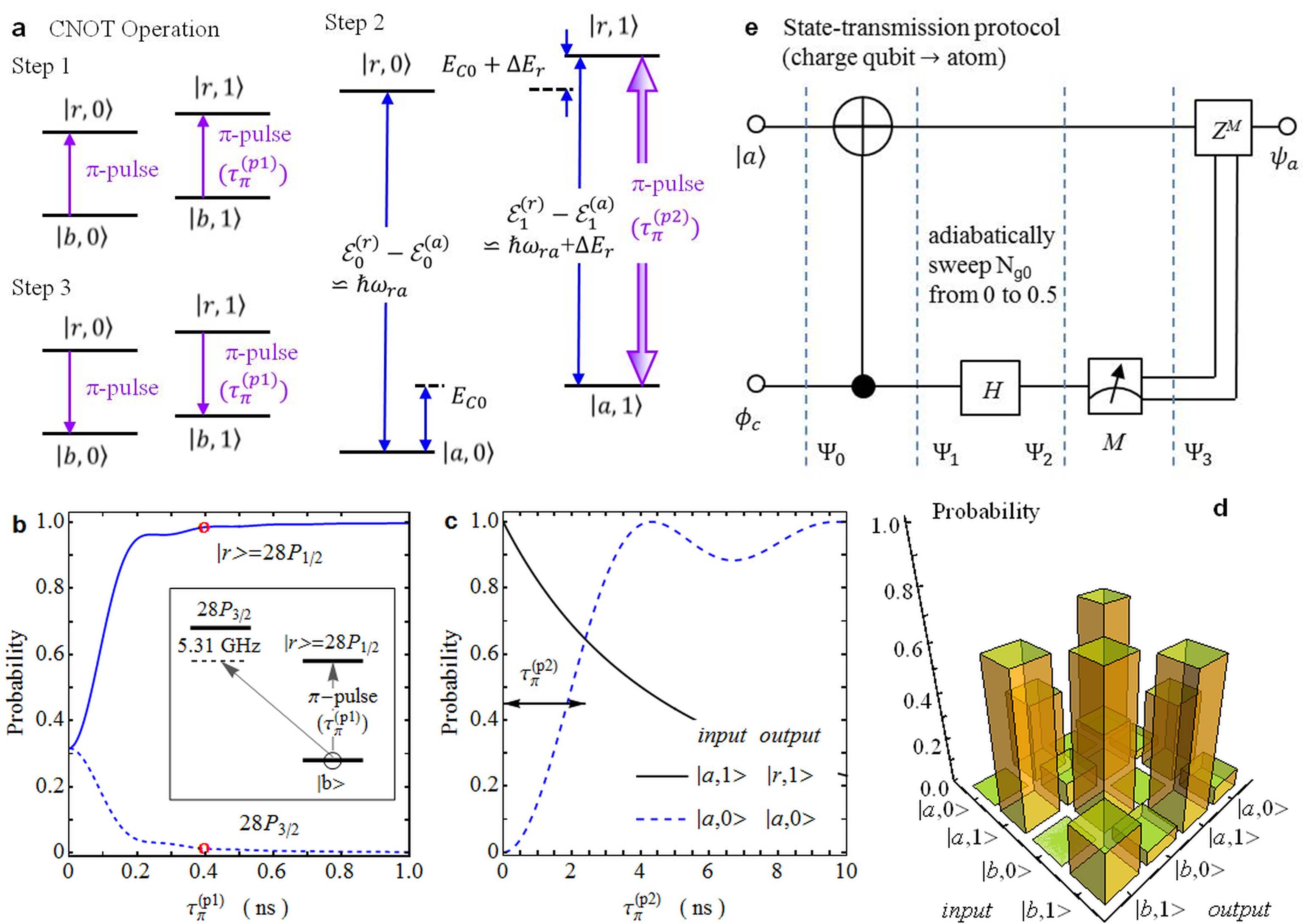}\\
\caption{{\bf Quantum-state transmission from charge qubit to atom.} (a) Scheme of two-qubit CNOT operation, where the atom acts as the target qubit while the SC circuit plays the control role. $N_{g0}$ is set to be zero. In Step 1, the atomic population in $|b\rangle$ is completely excited to the Rydberg $|r\rangle$ state via the $\pi$-light pulses at 297 nm with the pulse length $\tau^{(p1)}_{\pi}$. Then, a $\pi$-light pulse with a duration $\tau^{(p2)}_{\pi}$ resonantly couples the $|a,1\rangle-|r,1\rangle$ transition to flip the atom between $|a\rangle$ and $|r\rangle$ in Step 2. Finally, the atom in $|r\rangle$ is mapped back onto $|b\rangle$ via the $\pi$-light pulses. (b) Effect of $28P_{3/2}$ on the atomic excitation from $|b\rangle$ to $|r\rangle=28P_{1/2}$. For the three-level system composed of $|b\rangle$, $|r\rangle$, and $28P_{3/2}$, the $\pi$-light pulse is resonantly coupled to the $|b\rangle-|r\rangle$ transition. The atom population in $|r\rangle$ reaches 0.98 at $\tau^{(p1)}_{\pi}=0.4$ ns. (c) Probabilities of the system in the $output$ state with an initial $input$ state as a function of the $\pi$-light pulse duration $\tau^{(p2)}_{\pi}$. (d) The register populations after the CNOT operation with different $input$ and $output$ states. The resulting process fidelity is ${\cal{F}}=0.64$. (e) State transmission protocol for transferring an arbitrary charge-qubit state $\psi_{c}=\mu|a\rangle+\nu|b\rangle$ to the atom which is initially prepared in $|a\rangle$. The single-qubit Hadamard gate $H$ and measurement $M$ act on the charge qubit and the phase-flip gate $Z$ is performed on the atom. After the implementation, the final atomic-qubit state is $\phi_{a}=\mu|a\rangle+\nu|b\rangle$.}\label{Figure-3}
\end{figure*}

The two-qubit CNOT gate, where the charge qubit plays the control role while the atom acts as the target qubit, can be simply implemented via three steps (see Fig.~\ref{Figure-3}a): (1) Two $\pi$-light pulses with the pulse duration $\tau^{(p1)}_{\pi}$ are applied resonantly on the $|b,0\rangle-|r,0\rangle$ and $|b,1\rangle-|r,1\rangle$ transitions to transfer the atomic population in $|b\rangle$ completely to $|r\rangle$. (2) A $\pi$-light pulse with the pulse length $\tau^{(p2)}_{\pi}$ is employed to resonantly couple to the $|a,1\rangle-|r,1\rangle$ transition. The atomic state is flipped between $|a\rangle$ and $|r\rangle$ when the charge qubit is in $|1\rangle$. By contrast, the populations in $|a,0\rangle$ and $|r,0\rangle$ are weakly affected. (3) Two $\pi$-light pulses with the duration $\tau^{(p1)}_{\pi}$ are applied again to map the atomic component in $|r\rangle$ back onto $|b\rangle$. The extra phase acquired in the gate operation can be canceled by the local operations on the atom~\cite{PRL:Jaksch2000}.

The fine-structure splitting between $|r\rangle=28P_{1/2}$ and $28P_{3/2}$ is $2\pi\times5.31$ GHz~\cite{PRA:Li2003}. To suppress the influence of $28P_{3/2}$ on the atom transfer between $|b\rangle$ and $|r\rangle$ in steps (1) and (3), the $\pi$-light pulse duration is chosen to be $\tau^{(p1)}_{\pi}=0.4$ ns, much shorter than the decoherence time of the SC circuit, with the corresponding Rabi frequency of $2\pi\times1.25$ GHz (see Fig.~\ref{Figure-3}b). In step (2), the limited frequency difference between two $|a,0\rangle-|r,0\rangle$ and $|a,1\rangle-|r,1\rangle$ transitions extends the $\pi$-pulse length $\tau^{(p2)}_{\pi}$ and, hence, the the relaxation and dephasing of charge qubit reduces the fidelity of two-qubit logic gate as shown in Fig.~\ref{Figure-3}b. We set $\tau^{(p2)}_{\pi}=2.4$ ns to obtain the optimal CNOT truth table (see Fig.~\ref{Figure-3}c). The total gate duration is $2\tau^{(p1)}_{\pi}+\tau^{(p2)}_{\pi}=3.2$ ns and the resulting process fidelity is ${\cal{F}}=0.64$.

After performing the CNOT gate, one can transmit an arbitrary charge-qubit state $\phi_{c}=\mu|0\rangle+\nu|1\rangle$ to the atom via the following three steps (see Fig.~\ref{Figure-3}c): (1) CNOT operation: The hybrid system is initially prepared in $\Psi_{0}=|a\rangle\bigotimes(\mu|0\rangle+\nu|1\rangle)=\mu|a,0\rangle+\nu|a,1\rangle$ at $N_{g0}=0$. After the CNOT gate, we arrive at the system state $\Psi_{1}=\mu|a,0\rangle+\nu|b,1\rangle$. (2) Hadamard transform: The offset charge $N_{g0}$ is increased to 0.5 adiabatically. Two components $|a,0\rangle$ and $|b,1\rangle$ in $\Psi_{1}$ follow the adiabatic energy bands ${\mathcal{E}}^{(a)}_{0}$ and ${\mathcal{E}}^{(b)}_{1}$, respectively. Actually, it is unnecessary to maintain the sweep rate of $N_{g0}$ constantly from 0 to 0.5. Initially, $N_{g0}$ adiabatically raises from 0 at a large rate. When $N_{g0}$ approaches to 0.5, the sweep rate decreases to a relative low value. After the local operations for canceling the extra accumulated phases~\cite{JPA:Obada2012}, the system state becomes $\Psi_{2}=\mu|a,+\rangle+\nu|b,-\rangle=\frac{1}{\sqrt{2}}(\mu|a\rangle+\nu|b\rangle)\bigotimes|0\rangle+\frac{1}{\sqrt{2}}(\mu|a\rangle-\nu|b\rangle)\bigotimes|1\rangle$. (3) Projective measurement: We measure the excess Cooper pairs in the box. If the readout is 0, we conclude the system state $\Psi_{3}=(\mu|a\rangle+\nu|b\rangle)\bigotimes|0\rangle$, otherwise, $\Psi_{3}=(\mu|a\rangle-\nu|b\rangle)\bigotimes|1\rangle$. Then, the offset charge $N_{g0}$ is reduced back to 0 rapidly. After an extra Pauli-$Z$ operation performed on the atom~\cite{PRL:Jaksch2000}, we finally obtain the atomic-qubit state $\phi_{a}=\mu|a\rangle+\nu|b\rangle$. Thus, the quantum-state transmission is accomplished.\\

{\bf DISCUSSION}\\

The SC circuits operate much faster than the atomic systems. Transmitting the atomic-qubit state to the SC circuit allows the rapid quantum gate operations. Nevertheless, these solid-state devices lose the coherence on a short time scale compared with the atomic systems. Transmitting the quantum state from the SC qubits to the atoms allows a long-time storage. To achieve this reversible state-transmission, we have proposed a hybrid structure composed of a charge qubit and an atomic qubit. Placing the atom inside the gate capacitor results in the atomic-state-dependence energy bands of charge qubit and the charge-state-dependence DC Stark shifts of atomic-qubit states. Applying the standard spectroscopy techniques and sweeping the gate charge bias (gate voltage) enable the quantum-state transmission between two different qubits as well as the universal two-qubit quantum gates.

As is known, the strong coupling to the local electromagnetic environment leads to the short relaxation ($T_{1}$) and dephasing ($T_{2}<T_{1}$) times of SC circuits. For the common charge qubit discussed in this paper, the excess Cooper pairs in the box lose their coherence after about 10 ns~\cite{PRL:Nakamura2002,PRB:Duty2004,PRA:Koch2007}. Based on our physical specification, the state transmission can be accomplished within the coherence time of common charge qubit. Nonetheless, the relaxation effects of charge qubit still limit the fidelity of two-qubit gate operations.

In our hybrid system, the atom locates close to the gate-capacitor plates. The inhomogeneous Stark effect from the adsorbate electric fields on the cryogenic surface imposes a severe limitation to the coherence of Rydberg states, reducing the fidelity of state transmission~\cite{PRL:Chan2014}. However, the effects of stray fields might be circumvented via coating the surfaces with adsorbates~\cite{RPA:HermannAvigliano2014}. Moreover, measuring the distribution of stray fields above the chip surface based on Rydberg-electromagnetically-induced transparency~\cite{PRA:Thiele2015} provides a potential of canceling the uniform electric fields by offset fields. We expect the resulting coherence time of Rydberg atom much longer than that of charge qubit.

Our scheme for quantum state transmission in a superconducting charge qubit-atom hybrid opens a new prospect for quantum information processing, where the macroscopic SC devices rapidly process the quantum information which can be saved in the long-term storage composed of a microscopic atomic system. The protocols established in this paper also allow the information transfer between two distant SC qubits. After transmitting the quantum state of a SC qubit to a local atom, the quantum information encoded in this atom can be further transferred to another remote atom via a traveling qubit (photon)~\cite{Nature:Moehring2007,Science:Hofmann2012,Nature:Bernien2013}. Subsequently, the quantum state is transmitted to another distant SC interacting with the remote atom. Moreover, transmitting the quantum states of two entangled atoms to two distant SC qubits, respectively, results in a pair of remotely-entangled SC qubits.\\

{\bf Acknowledgments:} This research is supported by the National Research Foundation Singapore under its Competitive Research Programme (CRP Award No. NRF-CRP12-2013-03) and the Centre for Quantum Technologies, Singapore.

{\bf Author Contributions:} R.D. envisaged the concept of physical model. D.Y. did the calculation and analysis. D.Y. and M.M.V. provided the parameters of suitable Rydberg states. C.H. contributed the experimental realizable parameters. L.C.K., L.A., D.Y., and R.D. contributed the conceptual approach for analyzing the presented system. All authors participated in discussions and writing of the text.

{\bf Competing financial interests:} The authors declare no competing financial interest.

\begin{table}
\centering
\begin{tabular}{l l l l l}
{\bf Charge-Qubit Structure} & & \\
\hline
Physical parameters~~~~~~~~~~~~~~~~~~~~~~~~~~&Symbol~~~~~~~& Value \\
\hline
Self-capacitance of JJ (aF) & $C_{j}$ & 30 \\
Josephson coupling energy (GHz) & $E_{J}/(2\pi\hbar)$ & 1.2 \\
Critical current (nA) & $I_{c}$ & 2.4 \\
Empty gate capacitance (aF) & $C_{g0}$ & 265.6 \\
Plate area ($\mu$m$^{2}$) & $s$ & $6\times6$ \\
Interplate distance ($\mu$m) & $l$ & 1.2 \\
Empty charging energy (GHz) & $E_{C0}/(2\pi\hbar)$ & 262.1 \\
Electric field amplitude (V/cm) & ${\mathpzc{E}}_{0}$ & 9.0 \\
Relaxation time (ns) & $T_{1}$ & 100 \\
Dephasing time (ns) & $T_{2}$ & 10\\
\hline
 & & \\
\multicolumn{3}{l}{{\bf Atomic Parameters} of $|r\rangle=nP_{1/2}$} \\
\hline
Physical parameters & Value & Value \\
\hline
Principle quantum number $n$ & 43 & 28 \\
Polarizability $\alpha_{r}$ [MHz/(V/cm)$^{2}$] & 98.3 & 4.0 \\
Ratio $\eta_{r}$ & 0.015 & 0 \\
Lifetime ($\mu$s) & 159.3 & 39.5 \\
Orbit diameter ($a_{0}$) & 3255.43 & 1284.7 \\
\hline
& & \\
\end{tabular}
\caption{{\bf Specifications of CPB structure and atomic parameters.} The geometry of gate capacitor should be designed carefully to enhance the ratio $\eta_{r}$ for $|r\rangle$ as much as possible. Then, the JJ self-capacitance $C_{j}$ is selected accordingly so that the ratio $E_{J}/E_{C0}$ is smaller than the charge-degenerate-spot separation $\Delta N_{g0}$, i.e., the system energy spectra with the atom in the hyperfine ground $|u=a,b\rangle$ and Rydberg $|r\rangle$ states can be distinguished. The relaxation and dephasing times of charge qubit are derived from~\cite{PRL:Nakamura2002,PRB:Duty2004,PRA:Koch2007}. The internal electric field amplitude ${\mathpzc{E}}_{r}$ needs to be weaker than the first DC Stark shift-induced avoided crossing field of $|r\rangle$~\cite{PRA:Sullivan1985}. The static polarizability $\alpha_{r}$ is derived from~\cite{Thesis:Pritchard}. Moreover, the orbit diameters (units of Bohr radius $a_{0}$) of the atom in $|r\rangle$ estimated from~\cite{JPB:Low2012} needs to be smaller than the interplate distance $l$. The lifetime of $|r\rangle$ in the cryogenic environment is calculated from~\cite{JPB:Branden2010}. For our specification of CPB structure, the charge-degenerate spot with the atom in $|r\rangle=43P_{1/2}$ is shifted from $N_{g0}=0.5$ by $\Delta N_{g0}=0.016$ (see Fig.~\ref{Figure-2}b), which is larger than the ratio $E_{J}/E_{C0}=0.004$. In comparison, the atom in $|r\rangle=28P_{1/2}$ hardly affects the behavior of excess Cooper pairs in the box because of $\eta_{r}\simeq0$ and the resulting DC Stark shift is $\Delta E_{r}=2\pi\hbar\times0.2$ GHz.}\label{Table}
\end{table}

\end{document}